\documentclass[sort&compress
  ,final            
  ]
  {aipproc}

\layoutstyle{8x11single}
\usepackage{enumerate}

\begin{document}
\newcommand{\be}{\begin{equation}}
\newcommand{\ee}{\end{equation}}
\newcommand{\bea}{\begin{eqnarray}}
\newcommand{\eea}{\end{eqnarray}}
\newcommand{\f}{\frac}  
\newcommand{\la}{\lambda}
\newcommand{\ve}{\varepsilon}
\newcommand{\ep}{\epsilon}
\newcommand{\da}{\downarrow}
\newcommand{\csb}{{\rm{SB}}}
\newcommand{\up}{\uparrow}
\newcommand{\V}{{\cal V}}
\newcommand{\ovl}{\overline}
\newcommand{\Ga}{\Gamma}
\newcommand{\ga}{\gamma}
\newcommand{\bra}{\langle}
\newcommand{\ket}{\rangle}
\newcommand{\eff}{_{\rm{eff}}}
\newcommand{\av}{{\rm{av}}}
\newcommand{\fl}{{\rm{fl}}}
\newcommand{\ina}{{\rm{in}}}
\newcommand{\wt}{\widetilde}
\newcommand{\ov}{\overline}
\newcommand{\G}{{\cal G}}
\newcommand{\Ha}{{\cal H}}
\newcommand{\sig}{\sigma}
\renewcommand{\le}{\leqslant}

\title{Multi-level and two-level models of the decay out of
  superdeformed bands }


\author{A.~J.~Sargeant}{
  address={Instituto de F\'isica, Universidade de S{\~{a}}o Paulo,
Caixa Postal 66318, 05315-970  S{\~{a}}o Paulo, SP, Brazil}
}


\author{M.~S.~Hussein}{
  address={Instituto de F\'isica, Universidade de S{\~{a}}o Paulo,
Caixa Postal 66318, 05315-970  S{\~{a}}o Paulo, SP, Brazil}
}


\author{A.~N.~Wilson}{
  address={Department of Nuclear Physics,
    Research School of Physical Sciences and Engineering,
    Australian National University, Canberra, ACT 0200 Australia}, 
  altaddress={Department of Physics and Theoretical Physics, 
    Faculty of Science, Australian National University,
    Canberra, ACT 0200 Australia}
}

\begin{abstract}
We compare a multi-level statistical model with a two-level model for the decay
out of superdeformed rotational bands in atomic nuclei. 
We conclude that while the 
models depend on different dimensionless combinations of the input parameters
and differ in certain limits, they essentially 
agree in the cases where experimental data is currently available. The 
implications of this conclusion are discussed.
\end{abstract}
\maketitle


\section{Introduction}
Superdeformed (SD) rotational bands occur in the second minimum 
of the potential energy surface in deformation space.
They have been observed in several  
mass regions around $A$=20, 40, 80, 130, 150, 165, 190 and 240
\cite{Singh:2002}. Characteristic of the decay of most superdeformed bands
is the sudden disappearance of the total intra-band decay intensity when a
certain spin is reached. The intensity reappears as
electromagnetic transitions in the normal deformed (ND) minimum. 
In certain special cases the decay path subsequent to the
SD band is known completely as it is for $^{133}$Nd 
\cite{Bazzacco:1994}, or almost completely as for $^{59}$Cu
\cite{Andreoiu:2003}. 
More typical however is the situation in the $A$ = 80, 150 and 190 regions,
where the decay from superdeformed to normal states is spread over
many different available paths, making observation of discrete $\gamma$
rays linking SD and ND states very difficult.  Because of this
fragmentation of the SD intensity, experimentalists have only been able
to identify a small number of the strongest paths in a few cases
\cite{Chiara:2005,Lauritsen:2002,Lopezmartens:1996,Hauschild:1997,Khoo:1996,%
Hackman:1997,Wilson:2003,Siem:2004} 
and these only when very
large data sets have been obtained.
In these more typical 
regions it is reasonably clear that there remains barrier between 
the minima at the decay-out spin and consequently that the decay-out occurs 
by tunneling 
\cite{Shimizu:1992,Shimizu:1993,Shimizu:2001,Yoshida:2001,Lagergren:2004}.

The fact that the decay path is unknown has been thought to indicate
that a statistical treatment of the ND states and their coupling 
to the SD band 
\cite{Vigezzi:1990a,Vigezzi:1990b} is appropriate. 
This approach is expected be valid when the ND level density is so high  
as to make a useful description of the coupling to individual
ND states unfeasible. However, given that there is uncertainty concerning
the ND level density around the SD band there remains the 
possibility that the decay-out occurs entirely through a single ND level. 
The ND level density depends
strongly on the excitation energy of the SD state; as SD bands have been
observed at relatively low energies in the Pb isotopes, 
the point of view encompassed by a two-level model where the 
decaying SD state couples to a single ND state (itself decaying) could 
be more appropriate. Indeed, in 
Ref.~\cite{Adamian:2004} the decay-out in the $A$=190 region is attributed
to the crossing of the SD band with the nearest neighboring excited ND band.

In this contribution we propose to compare the two limiting 
descriptions mentioned in the previous paragraph as embodied by the 
multi-level statistical model of
Refs.~\cite{Gu:1999bv,Sargeant:2002sv} which exploit analogies of the SD decay
with the theory of compound nucleus reactions \cite{Feshbach:1992} and the 
two-level model of 
Refs.~\cite{Stafford:1999gz,Cardamone:2002,Cardamone:2003}.
Neither approach provides a microscopic description of the decay-out such
those provided by the cranked Nilsson-Strutinsky model of 
Refs.~\cite{Shimizu:1992,Shimizu:1993,Shimizu:2001,Yoshida:2001},
the cluster model of Ref.~\cite{Adamian:2004} and the generator coordinate 
model of Ref.~\cite{Bonche:1990}.
Rather, both models considered here are phenomenological models which attempt 
to provide simple formulas for the total intra-band decay intensity in terms 
of the most relevant parameters. 
Instead of analyzing the experimental data, in what
follows we shall only compare the statistical and two-level models
in their formal aspects.

\section{Multi-level statistical models}
\begin{center}
\begin{figure}
\includegraphics[height=0.33\textheight]{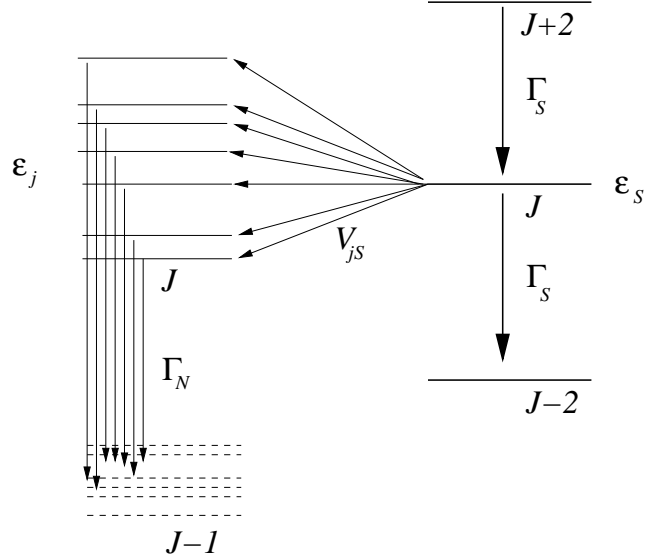}
\caption{Schematic diagram of the multi-level statistical model. The decaying
SD state under consideration has energy $\ve_S$ and spin $J$ and mixes via 
the coupling $V_{jS}$ with
ND states of the same spin whose energies, $\ve_j$, are the eigenvalues 
of the GOE Hamiltonian $H_{NN'}$. The SD (ND) states have a common 
electromagnetic width $\Ga_S$ ($\Ga_N$).}
\label{fig1}
\end{figure}
\end{center}
Let us denote the decaying SD state by $|S\ket$, its energy by $\ve_S$
and its electromagnetic width by $\Ga_S$. 
It is coupled to the ND states $|N\ket$, $N=1,...,K$ by
$V_{NS}$. The ND states are modeled by the GOE Hamiltonian $H_{NN'}$
and are assumed additionally to have a common electromagnetic width $\Ga_N$. 
The intra-band decay (see Fig. \ref{fig1}) is described by the 
Green's function \cite{Weidenmuller:1998xf,Gu:1999bv}
\be
G_{SS}(E)=\f{1}{E-\ve_S+i\Ga_S/2
-\sum_{NN'}V_{SN}g_{NN'}V_{N'S}}; \hspace{1cm}
(g^{-1})_{NN'}=E\delta_{NN'}-H_{NN'}+\f{i}{2}\Ga_N\delta_{NN'}
\label{GSS}
\ee
and the decay-out by the Green's function
\be
G_{NS}(E)=\sum_{N'}(e^{-1})_{NN'}\f{H_{N'S}}{(E-\ve_S+\f{i}{2}\Ga_S)};
\hspace{1cm}
e_{NN'}=E\delta_{NN'}-H_{NN'}+\f{i}{2}\Ga_N\delta_{NN'}
-\f{H_{NS}H_{SN'}}{(E-\ve_S+\f{i}{2}\Ga_S)}.
\label{GNS}
\ee
The total intra-band decay intensity is then given by
\be
F_S=\f{\Ga_S}{2\pi}
\int_{-\infty}^{\infty}
dE|G_{SS}(E)|^2
\label{F_S}
\ee
and the total decay-out intensity by
\be
F_N=\f{\Ga_N}{2\pi}
\int_{-\infty}^{\infty}
dE\sum_N|G_{NS}(E)|^2.
\label{F_N}
\ee
The ensemble average is defined by 
\be
\ov{F_S}=\int F_S(H_{NN'})P(H_{NN'})d[H_{NN'}],
\label{GOEav}
\ee
where $P(H_{NN'})$ is the probability distribution of the GOE 
and $d[H_{NN'}]$ is the product of the differentials of all independent
matrix elements of $H_{NN'}$ \cite{Guhr:1998ve}. Gu and Weidenm\"uller 
calculated $\ov{F_S}$ \cite{Gu:1999bv}
by adapting the result for the ensemble average of the 
compound nucleus cross-section derived by Verbaarschot et al. 
\cite{Verbaarschot:1985jn} to obtain, in the limit $K\to\infty$, 
\be
\ov{F_S}=F_S^\av+\ov{F_S^\fl},
\label{2F_S}
\ee
where \cite{Weidenmuller:1998xf}
\be
F_S^\av=\f{1}{1+\Ga/\Ga_S}
\label{2F_Sav}
\ee
and \cite{Gu:1999bv}
\begin{eqnarray}
&
\overline{F_S^{\rm fl}} = \frac{1}{16\pi\Gamma_S}
\int_{-\infty}^{\infty} {\rm d}E \int_{0}^{\infty} {\rm d}\lambda_{1}
\int_{0}^{\infty} {\rm d}\lambda_{2} \int_{0}^{1} {\rm d}\lambda 
\frac{(1-\lambda) \lambda |\lambda_1 - \lambda_2|} {((1+\lambda_1) 
\lambda_1 (1+\lambda_2) \lambda_2)^{1/2} (\lambda + \lambda_1)^2
(\lambda + \lambda_2)^2} 
\exp [-\frac{\pi \Gamma_N}{D} (\lambda_{1}+\lambda_{2}+2\lambda) ]
&
\nonumber \\
&
\frac{1-T \lambda} {(1+T \lambda_{1})^{1/2} (1+T\lambda_{2})^{1/2}}
\bigl (\ |\overline{S(E)}|^2 \ T^2 \ \
(\frac{\lambda_1}{1+T \lambda_1} + \frac{\lambda_2}{1+T \lambda_2} +
\frac{2 \lambda}{1-T \lambda})^2 
+ 2 \ T^2  \ \ (\frac{\lambda_1 (1+\lambda_1)} {(1+T \lambda_1)^2} + 
\frac{\lambda_2 (1+\lambda_2)}{(1+T \lambda_2)^2} + \frac{2\lambda 
(1-\lambda)}{(1-T \lambda)^2}) \ \bigr ) \ ,
&
\label{trip}
\end{eqnarray}
with
\[
\overline{S(E)} = \frac{E - E_0 - i \Gamma_S/2 + 
i \Ga/2}
{E - E_0 + i \Gamma_S/2 + i \Ga/2} \ , 
\hspace{.5cm}
T        = 1 - |\overline{S(E)}|^2 \nonumber     
= \frac{\Gamma_S \Ga}
{(E - E_0)^2 + (\Gamma_S + \Ga)^2/4} \ ,
\]  
\be
\Ga=2\pi V^2/D \hspace{.5cm} \mbox{and} \hspace{.5cm} 
V^2=\f{1}{K}\sum_NV_{NS}^2.
\label{Ga}
\ee
The density of ND levels is denoted by $D$.
From Eq.~(\ref{trip}) Gu and Weidenm\"uller deduced the fit formula
\be
\ov{F_S^{\rm{fl}}}=\left[1-0.9139\left(\Ga_N/D\right)^{0.2172}
\right]
\exp\left\{-\f{\left[0.4343\ln\left(\Ga/\Ga_S\right)
-0.45\left(\Ga_N/D\right)^{-0.1303}\right]^2}
{\left(\Ga_N/D\right)^{-0.1477}}\right\}.
\label{gufit}
\ee

Energy averaging provides an alternative approach to calculating the
average values of observables which depend on a random Hamiltonian. The 
energy average of $G(E)$ is given by
\be
\ov{G(E)}=\int G(E') p(E,E')dE',
\ee 
where the smoothing function $p(E,E')$ is normally a Lorentzian or a box
function \cite{Feshbach:1992}. The energy average is carried out for 
a single realization of the GOE Hamiltonian $H_{NN'}$ and is expected to
be equal to the ensemble average to the extent that the GOE is ergodic.  
In practice calculations proceed by choosing a representation - the optical
background representation of Kawai, Kerman and McVoy \cite{Kawai:1973}
- which is defined such that the couplings $V_{NS}$ have statistical 
properties which are convenient for analytical calculation. 
(It was checked numerically in
Ref.~\cite{Dagdeviren:1985} that the properties of the $V_{NS}$ 
which are normally assumed do indeed obtain from the underlying random 
Hamiltonian.)  In order to obtain analytical results it is further necessary 
to assume that $\Ga_N/D\gg 1$ which is the principal limitation of the 
energy averaging technique. The optical background representation  
was used in Ref.~\cite{Sargeant:2002sv} to calculate $\ov{F_S}$. 
The result has the same form as Eq.~(\ref{2F_S}) with $F^\av_S$ still given
by Eq.~(\ref{2F_Sav}) but now
\be
\ov{F_S^\fl}=2\left(\pi\Ga_N/D\right)^{-1}
F_S^\av\left(1-F_S^\av\right)^2,\hspace{1cm}\Ga_N/D\gg 1.
\label{F_Sfl}
\ee

An advantage of the energy averaging technique is that it also
yields an analytical expression for the variance of the decay intensity 
\cite{Sargeant:2002sv}:
\bea\nonumber
&&\ov{\left(\Delta F_S \right)^2}
=\ov{\left(F_S-\ov{F_S}\right)^2}
\\
&&\qquad
=\left(\f{\Ga_S}{2\pi}\right)^2
\int_{-\infty}^{\infty}dE\int_{-\infty}^{\infty}dE'\left[\hspace{2mm}
\ov{\left|G_{SS}^{\rm{fl}}(E){G_{SS}^{\rm{fl}}(E')}^*\right|^2}\right.
+\left.2\mbox{Re}
\ov{{G_{SS}(E)}^*}\hspace{2mm}\ov{G_{SS}(E')}
\hspace{2mm}\ov{G_{SS}^{\rm{fl}}(E){G_{SS}^{\rm{fl}}(E')}^*}\right]
\nonumber
\\
&&\qquad
={\ov{F_S^\fl}}^2f_1\left(\xi\right)
+2F_S^\av\ov{F_S^\fl}f_2\left(\xi\right),
\label{var}
\eea
where $\xi=(\Ga_S+\Ga)/\Ga_N$
and
\be
f_1\left(\xi\right)=\f{1}{\left(1+\xi\right)}+\f{\xi}{\left(1+\xi\right)^2}
+\f{\xi^2}{2\left(1+\xi\right)^3}, \hspace{1cm}
f_2\left(\xi\right)=\f{1}{2\left(1+\xi\right)}.
\label{f1f2}
\ee

From Eqs.~(\ref{2F_S}), (\ref{2F_Sav}), (\ref{trip}) and (\ref{F_Sfl}) it is
seen that $\ov{F_S}$ depends only on the two dimensionless variables
$\Ga/\Ga_S$ and $\Ga_N/D$. However it is possible to construct three 
independent dimensionless variables from the input variables
$\Ga$, $\Ga_S$, $\Ga_N$ and $D$.
In Ref.~\cite{Gu:1999bv} it is conjectured
using supersymmetry arguments \cite{Verbaarschot:1985jn,Davis:1989}
that the entire probability distribution of $F_S$ depends only on $\Ga/\Ga_S$ 
and $\Ga_N/D$. In contradiction to this conjecture
the expression for the variance of $F_S$ derived using the energy averaging
technique, Eq.~(\ref{var}), depends on the additional dimensionless
variable $(\Ga_S+\Ga)/\Ga_N$.

At present the energy dependence of $|G_{SS}(E)|^2$ and $|G_{NS}(E)|^2$
cannot be resolved experimentally
and the focus of theory has been on the analysis
of $F_S$ or $F_N$. Normally, $\Ga_S$, $\Ga_N$ and $D$ are estimated
theoretically and a value of $V$ is extracted from the experimental $F_S$
\cite{Krucken:2001we}. An intrinsic problem of the statistical model in its 
application to the decay out of SD bands is that the 
variance of $F_S$ is large when $\Ga_N/D \ll 1$ reflecting the fact that 
in this limit the positions of the individual ND states relative to 
the SD state are important \cite{Gu:1999bv}. 
Since $\Ga_N/D \ll 1$ for the cases which have been studied 
experimentally the preceding argument suggests that the 
error in the extracted $V$ must also be large.
However, from Eqs.~(\ref{var}) and (\ref{f1f2}) it can be seem 
that the variance of $F_S$ is suppressed not only for $\Ga_N/D\gg 1$
but also for $(\Ga+\Ga_S)/\Ga_N\gg 1$. 
This is physically plausible since in the decay of an ND state 
to the SD state the larger the value of $\Ga+\Ga_S$ the less important 
the exact position of the SD state.
\begin{center}
\begin{figure}
\includegraphics[height=0.33\textheight]{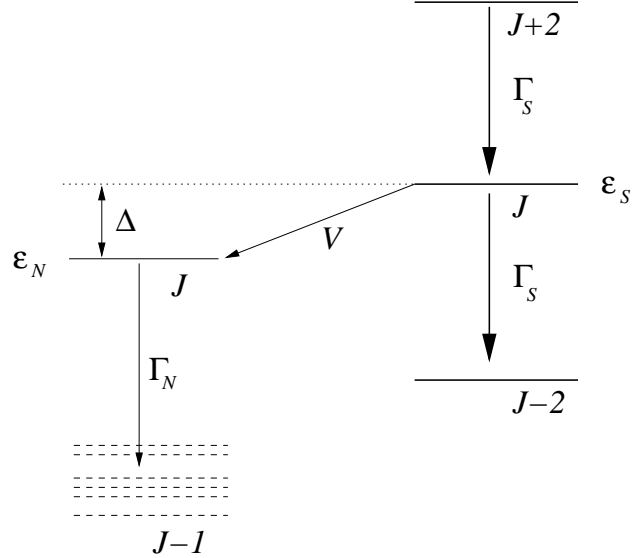}
\caption{Schematic diagram of the two-level mixing model.
The decaying SD state under consideration mixes via 
the coupling $V$ with a single ND state of the same spin and energy
$\ve_N=\ve_S+\Delta$.}
\label{fig2}
\end{figure}
\end{center}
\section{Two-level mixing model}
In the two-level mixing model the decaying SD state 
of energy $\ve_S$ and electromagnetic width $\Ga_S$ is coupled to
single ND state of energy $\ve_N$ and electromagnetic width $\Ga_N$
by the coupling matrix element $V$ (see Fig. \ref{fig2}).
In analogy to Eqs.~(\ref{GSS}) and (\ref{GNS}), the 
intra-band decay is described by the Green's function
\be
G_{SS}(E)=\f{1}{E-\ve_S+i\Ga_S/2-\f{V^2}{E-\ve_N+\f{i}{2}\Ga_N}}
\label{GSS2}
\ee
and the decay-out by the Green's function
\be
G_{NS}(E)=\f{V}{(E-\ve_N+\f{i}{2}\Ga_N)(E-\ve_S+\f{i}{2}\Ga_S)-V^2}.
\label{GNS2}
\ee
The decay-out intensity, $F_N$, is given by Eq.~(\ref{F_N}) (instead of a
sum over $N$ there is just one term) and an exact calculation 
yields \cite{Stafford:1999gz}
\be
F_N=\f{(1+\Ga_N/\Ga_S)V^2}{\Delta^2+\bar\Ga^2(1+4V^2/\Ga_N\Ga_S)},
\label{F_Nex}
\ee
where
\begin{equation}
\Gamma^\da = \f{2V^2\bar\Ga}{\Delta^2 + \bar\Ga^2},
\label{eqn:fermicsb}
\end{equation}
$\bar\Gamma=(\Ga_S+\Ga_N)/2$ and $\Delta=\ve_N-\ve_S$.
An alternative form is \cite{Cardamone:2003}
\be
F_N=\f{\Ga_N\Ga^\da/(\Ga_N+\Ga^\da)}
{\Ga_S+\Ga_N\Ga^\da/(\Ga_N+\Ga^\da)}.
\label{F_Ncard}
\ee
Since $F_S=1-F_N$ this implies that 
\be
F_S=(1+\Ga^\da/\Ga_S\f{\Ga_N}{\Ga^\da+\Ga_N})^{-1}.
\label{F_Sex}
\ee
From Eq.~(\ref{F_Nex}) it may be seen that $F_N$ depends on three
independent dimensionless variables which may be taken to be
$V/\Delta$, $\bar\Ga/\Delta$ and $\Ga_S/\Ga_N$. 
The calculations of Ref.~\cite{Adamian:2004} suggest that the value of 
$\Delta$ is rather important and in fact determines the decay-out spin
in the sense that the decay-out occurs very near to where the
SD band crosses the nearest excited ND band. 
The behavior of $\ov{F_S}$ as a function of 
$\Delta$ for a model related to the two-level model 
is discussed in Ref.~\cite{Sargeant:2004}. In the event that
$\Delta$ could be calculated it would be favorable to use
Eq.~(\ref{F_Nex}) to extract $V$ from the experimental $F_N$.

An alternative to calculating $\Delta$ is to assume that it obeys some 
probability distribution. In Ref.~\cite{Cardamone:2003} it was assumed
that $\Delta$ has the distribution 
\be
P(\Delta)=\int_0^\infty ds {\cal P}_s(\Delta)P(s),
\label{Pdel}
\ee
where
\be
P(s)=\f{\pi}{2}se^{-\pi s^2/4}\hspace{.5cm}\mbox{and}\hspace{.5cm}
{\cal P}_s(\Delta)=\f{1}{D}\Theta(\f{s}{2}-\f{|\Delta|}{D}).
\label{wigner}
\ee
In Ref.~\cite{Cardamone:2003} it was argued that the average
value $\bra|\Delta|\ket=D/4$ was the appropriate value to use
in the calculation of $F_N$. Alternatively, one could calculate 
the average of $F_N$ with respect to the distribution $P(\Delta)$:
\be
\bra F_N\ket=\int^{\infty}_{-\infty}F_N(\Delta)P(\Delta).
\label{F_Nav}
\ee
Since $\Delta=\ve_N-\ve_S$, Eq.~(\ref{F_Nav}) corresponds to an
average over the eigenvalues, $\ve_j$, of the GOE Hamiltonian 
of the statistical model. Thus, while Eq.~(\ref{F_Nav}) is related to the  
ensemble average, Eq.~(\ref{GOEav}), it is not identical with it since 
Eq.~(\ref{GOEav}) is an average over the $H_{NN'}$, ie. an average over both
eigenvalues and eigenvectors. It is clear from Eq.~(\ref{F_Nav}) that 
$\bra F_N\ket$ depends on the dimensionless variables
$V/D$, $\bar\Ga/D$ and $\Ga_S/\Ga_N$. 
Both $F_N(\Delta=D/4)$ and 
$\bra F_N\ket$ are compared with $\ov{F_N}$ of the statistical model in
the next section.

\begin{center}
\begin{figure}
\includegraphics[width=\textwidth]{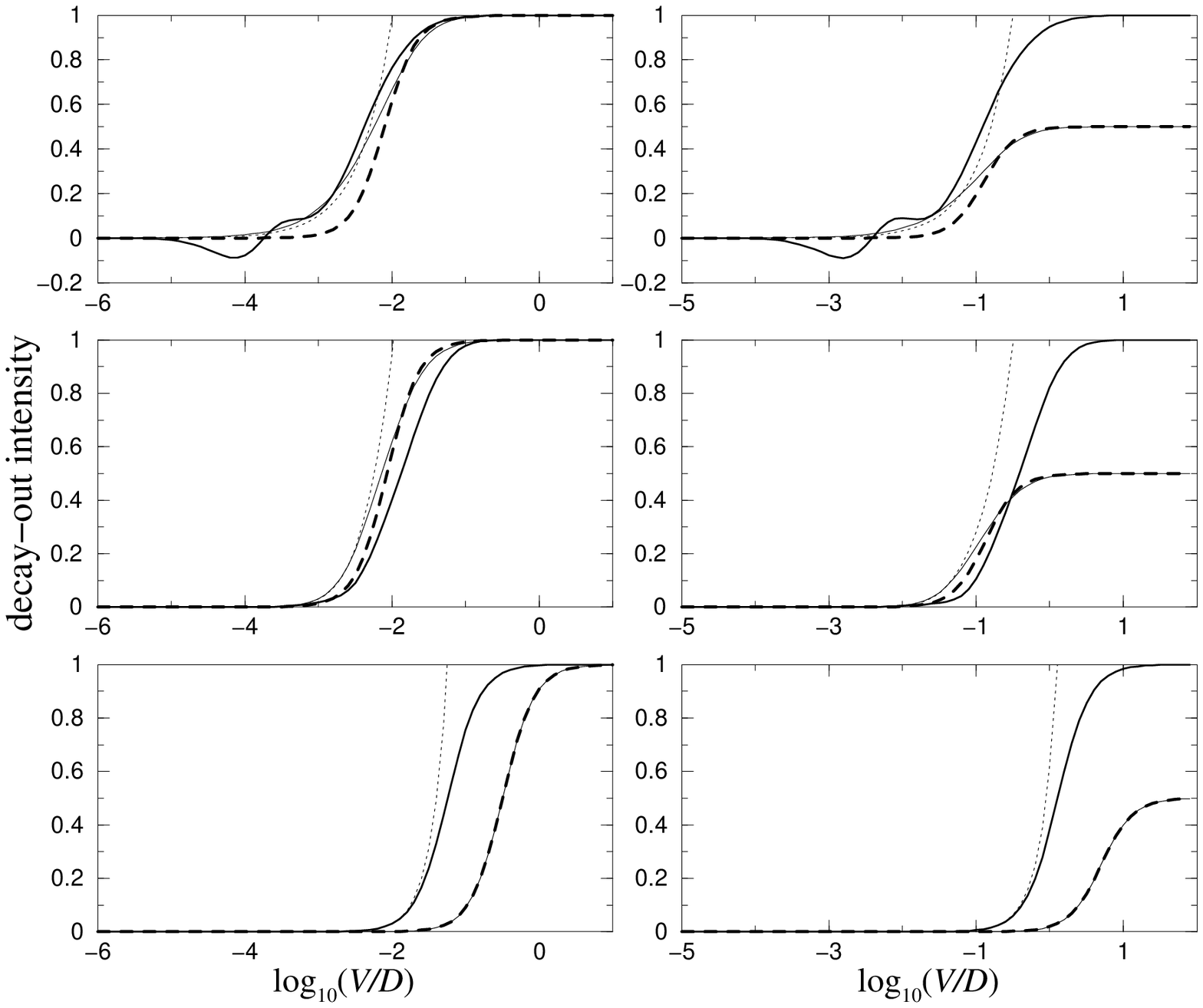}
\caption{The figure shows $1-\ov{F_S}$ (thick solid), $\bra F_N\ket$ 
(thin solid), $\bra F_N\ket_{\rm{U}}$ (dotted)
and $F_N(\Delta=D/4)$ (dashed) versus $\log_{10}(V/D)$, calculated using 
Eqs.~(\ref{2F_S}), (\ref{F_Nav}), (\ref{F_NavU}) and (\ref{F_Nex}) 
respectively. 
The left column has 
$\Ga_N/\Ga_S=1000$ and the right column has $\Ga_N/\Ga_S=1$.
The first, second and third rows have $\bar\Ga/D=10^{-4}$, 0.1 and 10.
For $\ov{F_S^\fl}$ we used Eq.~(\ref{gufit}) when $\bar\Ga/D=10^{-4}$, 0.1
and Eq.~(\ref{F_Sfl}) when $\bar\Ga/D=10$. The negative values of $F_N$ in
the first row are due to the fact that Eq.~(\ref{gufit}) is an approximation
to the exact result, Eq.~(\ref{trip}). Eq.~(\ref{gufit}) also appears to 
overestimate $\ov{F_S^\fl}$ in the middle row.
Note that $\bra F_N\ket$ 
and $F_N(\Delta=D/4)$ coincide in the bottom row.}
\label{fig3}
\end{figure}
\end{center}
\begin{center}
\begin{figure}
\includegraphics[width=\textwidth]{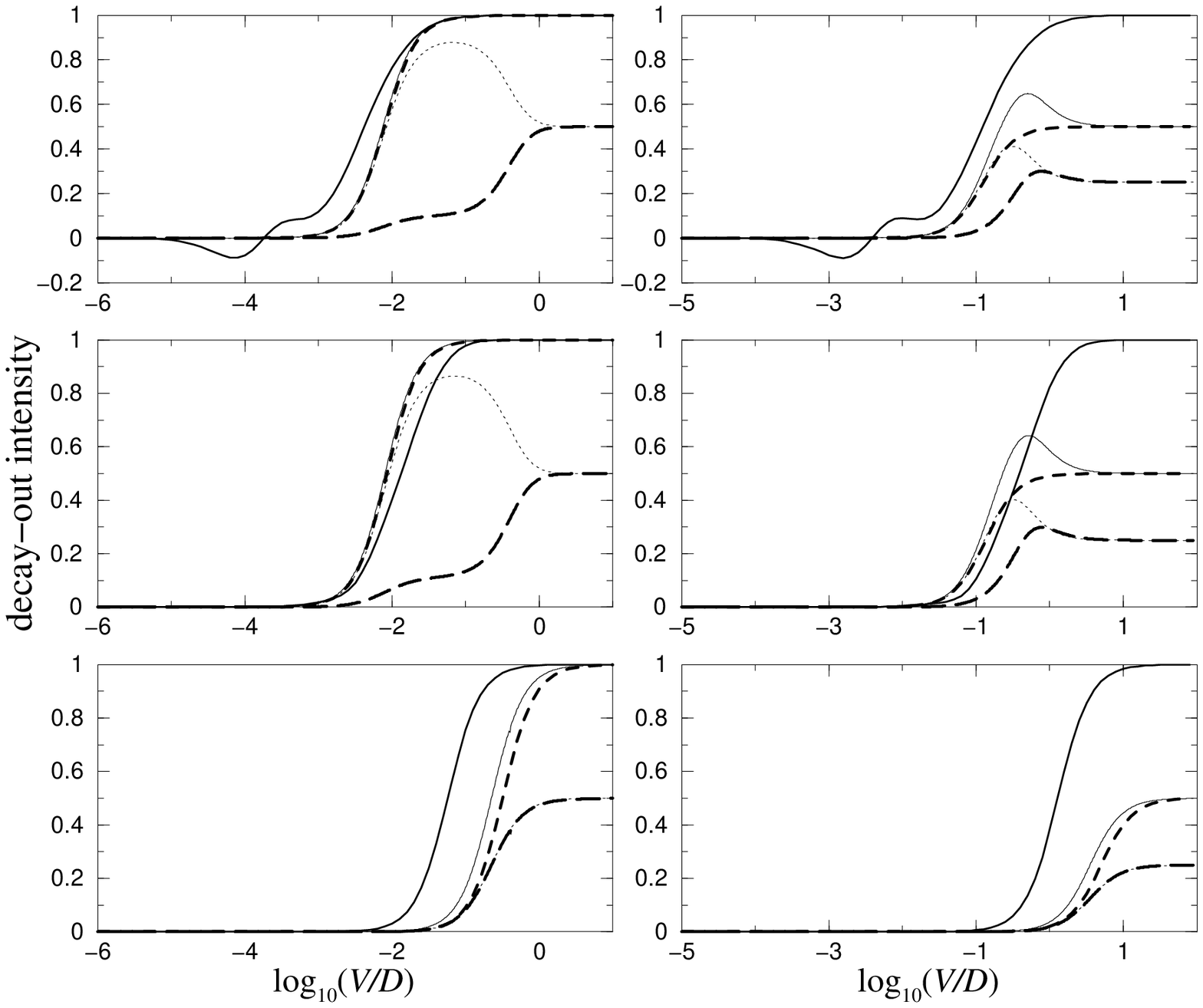}
\caption{Decay-out intensity for the three-level model. 
The dotted and long dashed curves are the branching ratios for decay 
to the nearest ND state and the next nearest ND state respectively while
the thin solid curve is the their sum (see Eq.~(\ref{FN3}) and the discussion
which follows it).
The thick solid and dashed curves are the same as in Fig.~\ref{fig3}
(they are the decay-out intensities for the multi-level statistical and
two-level models respectively) as are values of $\bar\Ga/D$ and $\Ga_N/\Ga_S$
used to plot the curves.
} 
\label{fig4}
\end{figure}
\end{center}
\section{Numerical comparison of the models}
The first thing which may be noted on comparing the statistical model and
the two-level model is that they depend on different dimensionless 
combinations of the
input parameters $V$, $D$, $\Ga_S$ and $\Ga_N$. The statistical model depends
on $\Ga/\Ga_S=2\pi(V/D)^2(D/\Ga_S)$ and $\Ga_N/D$ while the
two-level model depends on $V/D$, $\bar\Ga/D=(\Ga_N/D)(1+\Ga_S/\Ga_N)/2$ and
$\Ga_S/\Ga_N$. In addition, it may be deduced from Eq.~(\ref{F_Nex})
that $F_N$ in the two-level model increases monotonically from zero to 
$\Ga_N/(\Ga_S+\Ga_N)$ as $V$ is varied from zero to infinity  
whereas in the statistical model $F_N$ increases monotonically from zero
to 1.

A second difference occurs in the (physically not realized)
limit $\Ga_N/D\to\infty$. In this limit
$\ov{F_N}$ in the statistical model becomes independent of $\Ga_N$, being
given by $[1+\Ga_S/\Ga]^{-1}$, cf. Eq.~(\ref{2F_Sav}). The 
two-level model in the limit that $\bar\Ga\to \infty$ yields
$F_N=\Ga_N/(\Ga_N+\Ga_S)[1+\Ga_N\Ga_S/4V^2)]^{-1}$ which clearly does not
become independent of $\Ga_N$ as $\Ga_N/D\to \infty$.

It is useful to note that when a uniform distribution for $\Delta$ is
assumed then
\be
\bra F_N\ket_{\rm U}=\f{1}{D}\int^{\infty}_{-\infty}F_N(\Delta)d\Delta
=\f{2\pi V^2/D\Ga_S}{\sqrt{1+4V^2/\Ga_N\Ga_S}},
\label{F_NavU}
\ee
which reduces to $\Ga/\Ga_S$ in the limit that $V\ll \Ga_N,\Ga_S$. The
limit of small $V$ was considered for the statistical model in 
Ref.~\cite{Gu:1999bv} where a perturbation expansion of Eq.~(\ref{GNS})
was carried out and the ensemble average of $F_N$ calculated: the result
is also $\Ga/\Ga_S$.

In Fig.~\ref{fig3} where we plot $F_N$ versus $V/D$,
the considerations of the preceding paragraph are born out.
The graphs in the left hand column were calculated for 
$\Ga_N=1000\Ga_S$ and those in the right hand column for $\Ga_N=\Ga_S$.
The first, second and third rows were calculated for 
$\bar\Ga$/D=$10^{-4}$, 0.1 and 10 respectively. The dependence
on $\Ga_N/\Ga_S$ is clearly seen in the two-level model as is its absence in 
the statistical model. 
Eq.~(\ref{F_NavU}) is seen to agree with the statistical model for small 
$V$ but to be unphysical for large $V$. It can also be seen that 
$\bra F_N\ket$ and $F_N(\Delta=D/4)$ agree best with the statistical model
when $\bar\Ga/D$ is small. 
Indeed, when $\Ga_N\gg\Ga_S$ so that the coherence 
effects present in the two-level model are suppressed {\em and} $\bar\Ga\ll D$ 
then $\bra F_N\ket$ and $F_N(\Delta=D/4)$ are close to the statistical
model for all $V/D$. 
It is interesting that $\bra F_N\ket_{\rm U}$ agrees 
better with statistical model than $\bra F_N\ket$ when $\bar\Ga\gg 1$ suggesting
that in this region the average over the uniform distribution, 
Eq.~(\ref{F_NavU}), corresponds more closely to the ensemble average, 
Eq.~(\ref{GOEav}), than Eq.~(\ref{F_Nav}) does.
The case $\bar\Ga/D=10^{-4}$, $\Ga_N/\Ga_S=1000$ shown
in Fig. \ref{fig3} is representative of the mass 190 region. The considerations
in the present contribution thus go some way to explaining why the 
values of $V$ extracted \cite{Krucken:2001we} 
from experimental data in the mass 190 region
using the statistical model and those extracted using 
the two-level model were very similar.

In Ref.~\cite{Cardamone:2003} the validity of the two-level model was
investigated by calculating the decay-out intensity for a three-level
model where the decaying SD state is allowed to decay to the nearest 
neighbor and next-nearest-neighbor ND states. The decay-out intensity
may be written
\be
F^{(3)}_N=\sum_{j=1}^2F_j, \hspace{.5cm} F_j=\f{\Ga_N}{2\pi}
\int_{-\infty}^\infty dE|G_{jS}(E)|^2,
\label{FN3}
\ee
where $G_{jS}$ denotes the Green's function of Eq.~(\ref{GNS}) in the
basis of eigenvectors, $|j\ket$, of $H_{NN'}$. Clearly, $F^{(3)}_N$ depends 
on $\Delta_1$ and $\Delta_2$, the relative energies of the two ND states 
and on $V_1$ and $V_2$, their interactions with the SD state.
In Ref.~\cite{Cardamone:2003} a probability
distribution for $\Delta_2$ similar to Eq.~(\ref{Pdel}) was assumed
resulting in the average value $\bra\Delta_2\ket=3D/4$. 

In Fig.~\ref{fig4} we compare $\ov{F_N}$ of the multi-level statistical model 
and $F_N(\Delta=D/4)$ of the two-level model with $F^{(3)}_N$, $F_1$ and $F_2$ 
for $\Delta_1=D/4$, $\Delta_2=3D/4$ and $V_1=V_2=V$. In similarity to 
two-level case, the effects of coherence are pronounced 
in $F^{(3)}_N$ when $\Ga_N=\Ga_S$. 
It can be seen that $F^{(3)}_N$ moves in the direction of 
$\ov{F_N}$ relative to $F_N(\Delta=D/4)$. (As mentioned in 
the caption to Fig.~\ref{fig3}, $\ov{F_N}$ appears to be
underestimated in the middle row due
to use of the approximate Eq.~(\ref{gufit}).) 
We expect that $F^{(n)}_N$ calculated
analogously to $F^{(3)}_N$ would rapidly approach $\ov{F_N}$ with increasing
$n$. The fact that $F^{(3)}_N$ and $F_N(\Delta=D/4)$ are barely distinguishable
for the case $\bar\Ga/D=10^{-4}$ , $\Ga_N/\Ga_S=1000$ again suggests that the
two-level model is adequate to describe the SD decay in the mass 190 region.
It would be interesting to calculate the average of $F^{(3)}_N$ over $\Delta_1$
and $\Delta_2$ in analogy to Eqs.~(\ref{F_Nav}) and (\ref{F_NavU}).
\section{Conclusions}
We have compared a multi-level
statistical model and a two-level model of the decay
out of a superdeformed band with each other. We conclude that while the 
models depend on different dimensionless combinations of the input parameters
and yield different results for certain limiting cases, they essentially 
agree in the cases where experimental data are available. However, this
conclusion does not mean that we have reached a fully satisfactory description
of the decay-out of superdeformed bands. The two-level model
and multi-level statistical model level quite naturally only agree in the 
parameter regime where there are just two well defined levels interacting. 
However, this agreement, as it is described in the preceding section, 
surely has much to do with the use 
of the GOE probability distribution to calculate the ensemble 
average in Eq.~(\ref{GOEav}) and the use of the
Wigner nearest neighbor spacing distribution in 
obtaining Eq.~(\ref{Pdel}) for the probability distribution of $\Delta$.
The use of these two intimately related distributions to account for
the unknown parameters of the respective formulations means that
level fluctuations must be equally important in the two-level and 
multilevel statistical models. Consequently, the respective  
values of $V$ which can be extracted from the experimental $F_N$ using the
two models appear to be subject to similar errors. 
An analytical calculation of the variance of $F_N$, based on either model,
which is valid in the physically realized region would be useful in
assessing the errors.


\begin{theacknowledgments}
We thank D.~M.~Cardamone for providing us with the calculations for the
three-level model shown in Figure \ref{fig4} and for many comments on
a draft of this contribution. We benefited from discussions with DMC and
C.~A.~Stafford during the 
Nuclei and Mesoscopic Physics Workshop held at the NSCL
in October, 2004. AJS is also grateful to DMC for answering questions
about Ref.~\cite{Cardamone:2003} by email.
MSH is partly supported by FAPESP and the CNPq and both MSH and AJS are 
supported by the Instituto de Mil\^enio de 
Informa\c c\~ao Qu\^antica - MCT, all of Brazil.
\end{theacknowledgments}



\bibliographystyle{aipproc}   

\bibliography{sd,sargeant,books,rmt,compound}

\begin{thebibliography}{34}
\expandafter\ifx\csname natexlab\endcsname\relax\def\natexlab#1{#1}\fi
\providecommand{\enquote}[1]{``#1''}
\expandafter\ifx\csname url\endcsname\relax
  \def\url#1{\texttt{#1}}\fi
\expandafter\ifx\csname urlprefix\endcsname\relax\def\urlprefix{URL }\fi
\providecommand{\eprint}[2][]{\url{#2}}

\bibitem[{Singh} et~al.(2002)]{Singh:2002}
B.~{Singh}, R.~{Zywina}, and R.~B. {Firestone}, \emph{Nucl. Data Sheets},
  \textbf{97}, 241--592 (2002).

\bibitem[{Bazzacco} et~al.(1994)]{Bazzacco:1994}
D.~{Bazzacco}, F.~{Brandolini}, R.~{Burch}, S.~{Lunardi}, E.~{Maglione}, N.~H.
  {Medina}, P.~{Pavan}, C.~{Rossi-Alvarez}, G.~{de Angelis}, D.~{de Acuna},
  M.~{de Poli}, J.~{Rico}, D.~{Bucurescu}, and C.~{Ur}, \emph{Phys. Rev. C},
  \textbf{49}, 2281 (1994).

\bibitem[{Andreoiu} et~al.(2003)]{Andreoiu:2003}
C.~{Andreoiu}, T.~{D{\o}ssing}, C.~{Fahlander}, I.~{Ragnarsson}, D.~{Rudolph},
  S.~{Aberg}, R.~A. {Austin}, M.~P. {Carpenter}, R.~M. {Clark}, R.~V.
  {Janssens}, T.~L. {Khoo}, F.~G. {Kondev}, T.~{Lauritsen}, T.~{Rodinger},
  D.~G. {Sarantites}, D.~{Seweryniak}, T.~{Steinhardt}, C.~E. {Svensson},
  O.~{Thelen}, and J.~C. {Waddington}, \emph{Phys. Rev. Lett.}, \textbf{91},
  232502 (2003).

\bibitem[Chiara et~al.(2005)]{Chiara:2005}
C.~J. Chiara, et~al., \emph{AIP Conf. Proc.} (2005), proceedings of the
  Conference on Nuclei at the Limits, Argonne National Laboratory, July 2004
  (to be published).

\bibitem[{Lauritsen} et~al.(2002)]{Lauritsen:2002}
T.~{Lauritsen}, M.~P. {Carpenter}, T.~{D{\o}ssing}, P.~{Fallon}, B.~{Herskind},
  R.~V. {Janssens}, D.~G. {Jenkins}, T.~L. {Khoo}, F.~G. {Kondev},
  A.~{Lopez-Martens}, A.~O. {Macchiavelli}, D.~{Ward}, K.~S. {Abu Saleem},
  I.~{Ahmad}, R.~{Clark}, M.~{Cromaz}, J.~P. {Greene}, F.~{Hannachi}, A.~M.
  {Heinz}, A.~{Korichi}, G.~{Lane}, C.~J. {Lister}, P.~{Reiter},
  D.~{Seweryniak}, S.~{Siem}, R.~C. {Vondrasek}, and I.~{Wiedenh{\" o}ver},
  \emph{Phys. Rev. Lett.}, \textbf{88}, 042501 (2002).

\bibitem[{Lopez-Martens} et~al.(1996)]{Lopezmartens:1996}
A.~{Lopez-Martens}, F.~{Hannachi}, A.~{Korichi}, C.~{Sch{\" u}ck},
  E.~{Gueorguieva}, C.~{Vieu}, B.~{Haas}, R.~{Lucas}, A.~{Astier},
  G.~{Baldsiefen}, M.~{Carpenter}, G.~{de France}, R.~{Duffait}, L.~{Ducroux},
  Y.~{Le Coz}, C.~{Finck}, A.~{Gorgen}, H.~{H{\" u}bel}, T.~L. {Khoo},
  T.~{Lauritsen}, M.~{Meyer}, D.~{Pr{\' e}vost}, N.~{Redon}, C.~{Rigollet},
  H.~{Savajols}, J.~F. {Sharpey-Schafer}, O.~{Stezowski}, C.~{Theisen}, U.~V.
  {Severen}, J.~P. {Vivien}, and A.~N. {Wilson}, \emph{Phys. Lett. B},
  \textbf{380}, 18--23 (1996).

\bibitem[{Hauschild} et~al.(1997)]{Hauschild:1997}
K.~{Hauschild}, L.~A. {Bernstein}, J.~A. {Becker}, D.~E. {Archer}, R.~W.
  {Bauer}, D.~P. {McNabb}, J.~A. {Cizewski}, K.-Y. {Ding}, W.~{Younes},
  R.~{Kr{\" u}cken}, R.~M. {Diamond}, R.~M. {Clark}, P.~{Fallon}, I.-Y. {Lee},
  A.~O. {Macchiavelli}, R.~{MacLeod}, G.~J. {Schmid}, M.~A. {Deleplanque},
  F.~S. {Stephens}, and W.~H. {Kelly}, \emph{Phys. Rev. C}, \textbf{55},
  2819--2825 (1997).

\bibitem[{Khoo} et~al.(1996)]{Khoo:1996}
T.~L. {Khoo}, M.~P. {Carpenter}, T.~{Lauritsen}, D.~{Ackermann}, I.~{Ahmad},
  D.~J. {Blumenthal}, S.~M. {Fischer}, R.~V. {Janssens}, D.~{Nisius}, E.~F.
  {Moore}, A.~{Lopez-Martens}, T.~{D{\o}ssing}, R.~{Kruecken}, S.~J.
  {Asztalos}, J.~A. {Becker}, L.~{Bernstein}, R.~M. {Clark}, M.~A.
  {Deleplanque}, R.~M. {Diamond}, P.~{Fallon}, L.~P. {Farris}, F.~{Hannachi},
  E.~A. {Henry}, A.~{Korichi}, I.~Y. {Lee}, A.~O. {Macchiavelli}, and F.~S.
  {Stephens}, \emph{Phys. Rev. Lett.}, \textbf{76}, 1583--1586 (1996).

\bibitem[{Hackman} et~al.(1997)]{Hackman:1997}
G.~{Hackman}, T.~L. {Khoo}, M.~P. {Carpenter}, T.~{Lauritsen},
  A.~{Lopez-Martens}, I.~J. {Calderin}, R.~V. {Janssens}, D.~{Ackermann},
  I.~{Ahmad}, S.~{Agarwala}, D.~J. {Blumenthal}, S.~M. {Fischer}, D.~{Nisius},
  P.~{Reiter}, J.~{Young}, H.~{Amro}, E.~F. {Moore}, F.~{Hannachi},
  A.~{Korichi}, I.~Y. {Lee}, A.~O. {Macchiavelli}, T.~{D{\o}ssing}, and
  T.~{Nakatsukasa}, \emph{Phys. Rev. Lett.}, \textbf{79}, 4100--4103 (1997).

\bibitem[{Wilson} et~al.(2003)]{Wilson:2003}
A.~N. {Wilson}, G.~D. {Dracoulis}, A.~P. {Byrne}, P.~M. {Davidson}, G.~J.
  {Lane}, R.~M. {Clark}, P.~{Fallon}, A.~{G{\" o}rgen}, A.~O. {Macchiavelli},
  and D.~{Ward}, \emph{Phys. Rev. Lett.}, \textbf{90}, 142501 (2003).

\bibitem[{Siem} et~al.(2004)]{Siem:2004}
S.~{Siem}, P.~{Reiter}, T.~L. {Khoo}, T.~{Lauritsen}, P.-H. {Heenen}, M.~P.
  {Carpenter}, I.~{Ahmad}, H.~{Amro}, I.~J. {Calderin}, T.~{D{\o}ssing},
  T.~{Duguet}, S.~M. {Fischer}, U.~{Garg}, D.~{Gassmann}, G.~{Hackman},
  F.~{Hannachi}, K.~{Hauschild}, R.~V. {Janssens}, B.~{Kharraja}, A.~{Korichi},
  I.-Y. {Lee}, A.~{Lopez-Martens}, A.~O. {Macchiavelli}, E.~F. {Moore},
  D.~{Nisius}, and C.~{Sch{\" u}ck}, \emph{Phys. Rev. C}, \textbf{70}, 014303
  (2004).

\bibitem[{Shimizu} et~al.(1992)]{Shimizu:1992}
Y.~R. {Shimizu}, F.~{Barranco}, R.~A. {Broglia}, T.~{D{\o}ssing}, and
  E.~{Vigezzi}, \emph{Phys. Lett. B}, \textbf{274}, 253 (1992).

\bibitem[{Shimizu} et~al.(1993)]{Shimizu:1993}
Y.~R. {Shimizu}, E.~{Vigezzi}, T.~{D{\o}ssing}, and R.~A. {Broglia},
  \emph{Nucl. Phys.}, \textbf{A557}, 99c (1993).

\bibitem[{Shimizu} et~al.(2001)]{Shimizu:2001}
Y.~R. {Shimizu}, M.~{Matsuo}, and K.~{Yoshida}, \emph{Nucl. Phys.},
  \textbf{A682}, 464--469 (2001).

\bibitem[{Yoshida} et~al.(2001)]{Yoshida:2001}
K.~{Yoshida}, M.~{Matsuo}, and Y.~R. {Shimizu}, \emph{Nucl. Phys.},
  \textbf{A696}, 85--122 (2001).

\bibitem[Lagergren and Cederwall(2004)]{Lagergren:2004}
K.~Lagergren, and B.~Cederwall, \emph{Eur. Phys. J. A}, \textbf{21}, 175--177
  (2004).

\bibitem[{Vigezzi} et~al.(1990{\natexlab{a}})]{Vigezzi:1990a}
E.~{Vigezzi}, R.~A. {Broglia}, and T.~{D{\o}ssing}, \emph{Nucl. Phys.},
  \textbf{A520}, 179c (1990{\natexlab{a}}).

\bibitem[{Vigezzi} et~al.(1990{\natexlab{b}})]{Vigezzi:1990b}
E.~{Vigezzi}, R.~A. {Broglia}, and T.~{D{\o}ssing}, \emph{Phys. Lett. B},
  \textbf{249}, 163 (1990{\natexlab{b}}).

\bibitem[{Adamian} et~al.(2004)]{Adamian:2004}
G.~G. {Adamian}, N.~V. {Antonenko}, R.~V. {Jolos}, Y.~V. {Palchikov},
  W.~{Scheid}, and T.~M. {Shneidman}, \emph{Phys. Rev. C}, \textbf{69}, 054310
  (2004).

\bibitem[Gu and Weidenm{\"u}ller(1999)]{Gu:1999bv}
J.-z. Gu, and H.~A. Weidenm{\"u}ller, \emph{Nucl. Phys.}, \textbf{A660},
  197--215 (1999).

\bibitem[Sargeant et~al.(2002)]{Sargeant:2002sv}
A.~J. Sargeant, M.~S. Hussein, M.~P. Pato, and M.~Ueda, \emph{Phys. Rev. C},
  \textbf{66}, 064301 (2002).

\bibitem[Feshbach(1992)]{Feshbach:1992}
H.~Feshbach, \emph{Theoretical Nuclear Physics: Nuclear Reactions}, Wiley,
  N.Y., 1992.

\bibitem[Stafford and Barrett(1999)]{Stafford:1999gz}
C.~A. Stafford, and B.~R. Barrett, \emph{Phys. Rev.}, \textbf{C60}, 051305
  (1999).

\bibitem[Cardamone et~al.(2002)]{Cardamone:2002}
D.~M. Cardamone, C.~A. Stafford, and B.~R. Barrett, \emph{Phys. Stat. Sol.},
  \textbf{230}, 419--423 (2002).

\bibitem[Cardamone et~al.(2003)]{Cardamone:2003}
D.~M. Cardamone, C.~A. Stafford, and B.~R. Barrett, \emph{Phys. Rev. Lett.},
  \textbf{91}, 102502 (2003).

\bibitem[{Bonche} et~al.(1990)]{Bonche:1990}
P.~{Bonche}, J.~{Dobaczewski}, H.~{Flocard}, P.~H. {Heenen}, S.~J. {Krieger},
  J.~{Meyer}, and M.~S. {Weiss}, \emph{Nucl. Phys.}, \textbf{A519}, 509 (1990).

\bibitem[Weidenmuller et~al.(1998)]{Weidenmuller:1998xf}
H.~A. Weidenmuller, P.~von Brentano, and B.~R. Barrett, \emph{Phys. Rev.
  Lett.}, \textbf{81}, 3603--3606 (1998).

\bibitem[Guhr et~al.(1998)]{Guhr:1998ve}
T.~Guhr, A.~Muller-Groeling, and H.~A. Weidenmuller, \emph{Phys. Rep.},
  \textbf{299}, 189--425 (1998).

\bibitem[Verbaarschot et~al.(1985)]{Verbaarschot:1985jn}
J.~J.~M. Verbaarschot, H.~A. Weidenmuller, and M.~R. Zirnbauer, \emph{Phys.
  Rep.}, \textbf{129}, 367--438 (1985).

\bibitem[Kawai et~al.(1973)]{Kawai:1973}
M.~Kawai, A.~K. Kerman, and K.~W. McVoy, \emph{Ann. Phys. (N.Y.)}, \textbf{75},
  156--170 (1973).

\bibitem[Dagdeviren and Kerman(1985)]{Dagdeviren:1985}
N.~R. Dagdeviren, and A.~K. Kerman, \emph{Ann. Phys. (N.Y.)}, \textbf{163},
  199--211 (1985).

\bibitem[Davis and Boose(1989)]{Davis:1989}
E.~D. Davis, and D.~Boose, \emph{Z. Phys. A}, \textbf{332}, 427--441 (1989).

\bibitem[Krucken et~al.(2001)]{Krucken:2001we}
R.~Krucken, A.~Dewald, P.~von Brentano, and H.~A. Weidenmuller, \emph{Phys.
  Rev. C}, \textbf{64}, 064316 (2001).

\bibitem[Sargeant et~al.(2004)]{Sargeant:2004}
A.~J. Sargeant, M.~S. Hussein, M.~P. Pato, N.~Takigawa, and M.~Ueda,
  \emph{Phys. Rev. C}, \textbf{69}, 067301 (2004).

\end{thebibliography}

\IfFileExists{\jobname.bbl}{}
 {\typeout{}
  \typeout{******************************************}
  \typeout{** Please run "bibtex \jobname" to optain}
  \typeout{** the bibliography and then re-run LaTeX}
  \typeout{** twice to fix the references!}
  \typeout{******************************************}
  \typeout{}
 }


\end{document}